\PassOptionsToClass{reprint}{revtex4-1}
\documentclass[aps,pre,twoside,showkeys,showpacs,floatfix]{revtex4-1}

\newcommand\revdate{\today}

\usepackage[utf8]{inputenc}
\usepackage{hyperref}
\usepackage{amsmath}
\usepackage{graphicx}
\usepackage{xspace}

\newcommand\textalpha{\ensuremath{\alpha}\xspace}
\newcommand\textsim{\ensuremath{\sim}}
\newcommand\textleq{\ensuremath{\leq}}

\renewcommand\textdegree{\ensuremath{^\circ}\xspace}

\begin{document}
\title[Energetics of voltage sensitivity]{
  Voltage sensing in ion channels:\\
  Mesoscale simulations of biological devices
}


\author{Alexander Peyser}
\affiliation{Department of Physiology and Biophysics, University of
  Miami}%
\affiliation{Computational Biophysics, German
  Research School for Simulation Sciences, 52425 J\"ulich}%
\author{Wolfgang Nonner}%
\email{wnonner@med.miami.edu}%
\affiliation{Department of Physiology and Biophysics, University of
  Miami}%
\date{\revdate}

\begin{abstract}\noindent
  Electrical signaling via voltage-gated ion channels depends upon the
  function of a voltage sensor (VS), identified with the S1--S4 domain
  in voltage-gated K$^+$ channels. Here we investigate some energetic
  aspects of the sliding-helix model of the VS using simulations based
  on VS charges, linear dielectrics and whole-body motion. Model
  electrostatics in voltage-clamped boundary conditions are solved
  using a boundary element method. The statistical mechanical
  consequences of the electrostatic configurational energy are
  computed to gain insight into the sliding-helix mechanism and to
  predict experimentally measured ensemble properties such as gating
  charge displaced by an applied voltage. Those consequences and
  ensemble properties are investigated for two alternate S4
  configurations, \textalpha- and $3_{10}$-helical. Both forms of VS
  are found to have an inherent electrostatic stability. Maximal
  charge displacement is limited by geometry, specifically the range
  of movement where S4 charges and countercharges overlap in the
  region of weak dielectric. Charge displacement responds more steeply
  to voltage in the \textalpha-helical than in the 3$_{10}$-helical
  sensor.  This difference is due to differences on the order of 0.1
  eV in the landscapes of electrostatic energy. As a step toward
  integrating these VS models into a full-channel model, we include a
  hypothetical external load in the Hamiltonian of the system and
  analyze the energetic input-output relation of the VS.
\end{abstract}

\maketitle

\section{Introduction}

Electrical excitability of cells is possible because the movement of a
few charges can control the flow of many charges. This principle ---
amplification --- led \citet{hodgkin:1952:quant} to their theory of
the action potential in terms of electrically controlled membrane
conductances. Electrically controlled conductances have been localized
to channel proteins conducting Na$^+$, K$^+$ or Ca$^{2+}$ ions across
the cell membrane. Voltage-controlled ionic conductances and the
controlling intrinsic charge movements (`gating currents') of ion
channels have been studied experimentally (reviewed in
\citealp{hille:2001}). Electrophysiology has been complemented by
techniques measuring channel topology, channel structure and the
change in channel function (reviewed in
\citealp{gandhi:2002,catterall:2010,gonzalez:2012}). Together, these
perspectives provide detailed information on the `voltage sensor' (VS)
common to these channels and exemplified by the S1--S4 transmembrane
segments of \emph{Shaker}-type $\text{K}^+$ channels. Here we use
simulation to determine consequences of voltage sensor models that are
based on the `sliding-helix' hypothesis in which the charged S4
segment responds to changes of the membrane electrical field by
sliding in a canal formed between other transmembrane domains. This
hypothesis, first proposed by \citet{catterall:1986}, qualitatively
correlates a large body of experimental work (reviewed in
\citealp{gandhi:2002,catterall:2010,gonzalez:2012}).

The relaxation times of the voltage sensor in K$^+$ channels are of
the order of milliseconds and thus are highly averaged manifestations
of atomic motion in a condensed phase, where momentum is scattered in
a picosecond \citep{berry:2000}. Furthermore, site-directed
mutagenesis experiments have shown that only a limited number of
amino-acid residues of a voltage-dependent ion channel are
individually important for voltage sensitivity
\citep{schoppa:1992,seoh:1996,aggarwal:1996,gandhi:2002}.  On the
basis of these experimental facts, we describe the voltage sensor in a
mesoscale model in which atoms are not made explicit. Because
amino-acid residues with formally charged side chains in the S4 as
well as S2 and S3 transmembrane segments strongly determine VS
function \citep{papazian:1995}, we explicitly represent the charges of
these residues in the model. These charges are embedded in an
environment of spatially non-uniform electrical polarizability. We
distinguish three dielectric regions in the model (membrane lipid,
baths, and protein) and describe electrical polarizabilities by linear
dielectric coefficients that are uniform within each region.

The resulting model shares important elements with models previously
investigated by \citet{lecar:2003}, \citet{grabe:2004}, and
\citet{silva:2009}. We include fewer geometrical details than those
models because we wish to start from a minimalistic model; with that
approach, the importance of structural features can be discovered as
features are included or varied. Additionally, we use self-consistent
methods in solving our models.

Electrophysiological data on voltage sensor function are recorded by a
macroscopic setup that picks up charge movements of a large ensemble
of ion channels while imposing a controlled voltage across the cell
membrane (`voltage clamp' \cite{hodgkin-katz:1952,armstrong:1973}). To
simulate such an experiment, we encapsulate the microscopic model of a
VS with conductors clamped to imposed potentials from external sources
while monitoring charge displacement at these electrodes. The
electrostatics of the system composed of the VS model and the
electrode setup is solved self-consistently. The configuration space
of the model VS (S4 translation and rotation) is systematically
sampled to construct a partition function based on the electrostatic
configurational energy. Using this partition function, we compute
ensemble expectations of observable random variables (e.g., of gating
charge displaced at an applied voltage). In this way our theoretical
results meet two criteria for practical usefulness: they are
consistent solutions of the physics included in the model, and they
directly pertain to macroscopic experimental results.

This paper describes a building block toward creating a mesoscale
physical model for voltage-gated ion channels. Such a channel
comprises a gated central pore (through which ions flow when the pore
is gated open) surrounded in the lipid membrane by four voltage sensor
domains. Here we describe and test a simulation system for a single
voltage sensor. Using this setup, we simulate an `idle' voltage sensor
to study its reconfiguration and energetics as the voltage changes while
no external work is done. As a step toward understanding how a sensor
might interact with other parts of the channel, we then test how a
simulated `load' alters the sensor movement and how the work done on
the load depends on the sensed variable, the voltage. These simulations
are done for two structures of voltage sensor in which the S4 helix is
arranged either in the \textalpha conformation or the
$3_{10}$-conformation. These helix forms are currently discussed as
possible alternatives for S4 structure
\citep{long:2007,khalili-araghi:2010,bjelkmar:2009,schwaiger:2011}. Simulations
with these alternate structures reveal substantial consequences for
voltage sensing.

\section{Model and boundary conditions}

\begin{figure*}
  \includegraphics{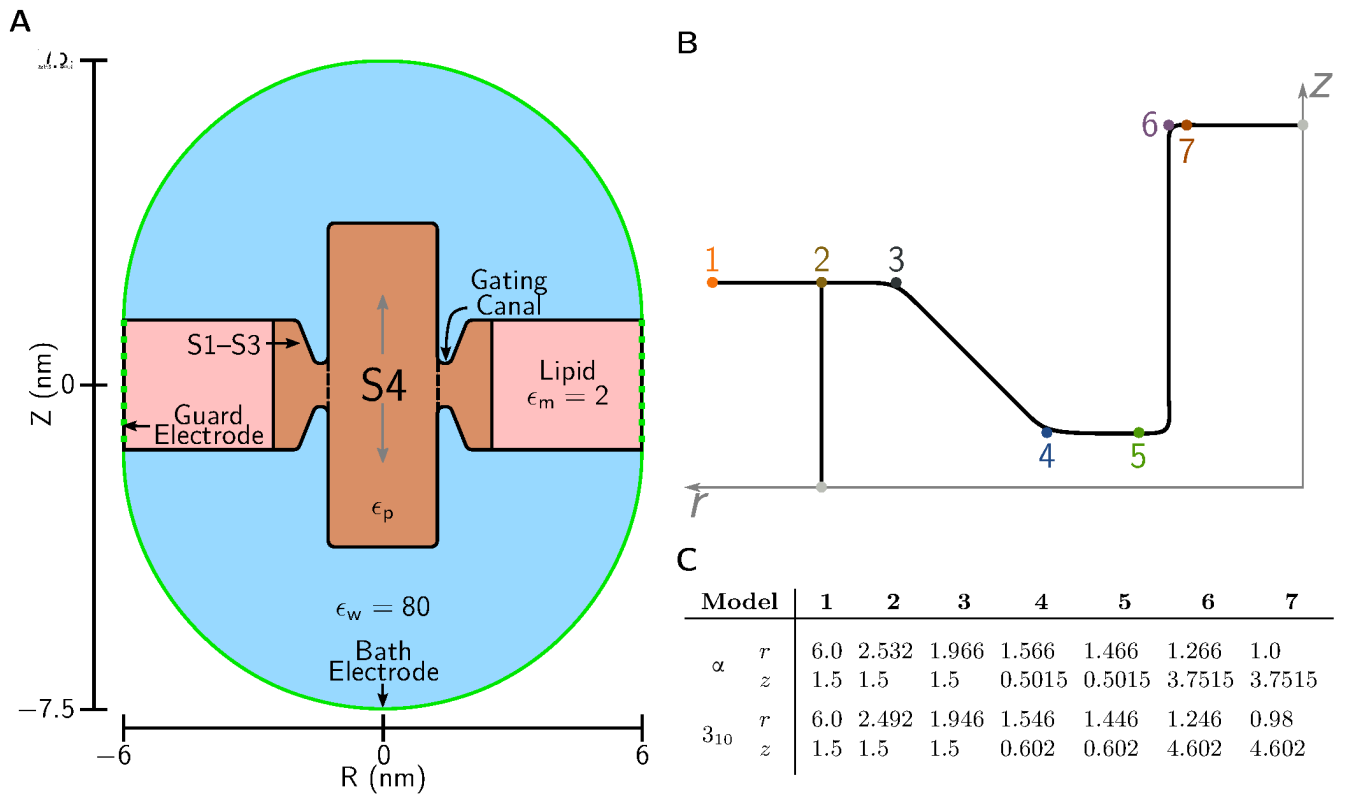}
  \caption{\emph{Simulation cell.} (A) The 3D setup is
    produced by rotating the cross-section about its vertical
    axis. \emph{Green lines} (labeled as bath \& guard electrodes) are
    electrode surfaces bounding the cell. \emph{Black lines} are
    dielectric boundaries separating uniform dielectric regions: baths
    (\emph{blue}, with dielectric coefficient
    $\epsilon_{\mathrm{w}}$), membrane lipids (\emph{pink}, with
    dielectric coefficient $\epsilon_{\mathrm{m}}$), and protein
    (\emph{brown} with dielectric coefficient
    $\epsilon_{\mathrm{p}}$). Charges of protein side chains
    (represented by colored balls in Fig.~\ref{fig:helix}) are
    embedded in the protein dielectric region in varied geometries.
    We thus simulate a single VS sensor domain (S1--S4) modeled as a
    central S4 cylinder surrounded by a ring of protein material
    including the S1--S3 transmembrane domains. The junction between
    these protein domains is narrowed to less than the membrane
    thickness by circular invaginations (`vestibules') leading up to
    the `gating canal' through which the S4 helix glides through the
    rest of the protein (dashed line).
    (B, C) \emph{Geometrical parameters of models.}
    (B) Mapping from geometrical positions in (A) to indexed
    geometrical parameters. Inflection points and lengths are varied
    among different models, depending on countercharge positions and
    helix conformation. Positions in $r$ (radial) and $z$ (axial)
    coordinates are marked by a colored point and an associated
    number. All corners are rounded with curvature radius of
    $0.15$~nm. Points 1,2,3,4, and 6 define the profile of the lipid
    and protein dielectrics, from the outermost end of the lipid
    domain (1) to the face of the S4 cylinder (6). Point 5 marks the
    radial position assigned to countercharges, point 7 the radial
    position of S4 charges.  (C) Coordinates (in nanometers) of the points
    defining membrane and protein metrics. Models are symmetrical with
    respect to the $z=0$ plane. See
    Figs.~\ref{fig:movie:first} \& \ref{fig:movie:last} for
    three-dimensional representations of these geometries.}
  \label{fig:cell}
\end{figure*}

Figure ~\ref{fig:cell} represents the simulation cell by an axial
cross-section of the radially symmetric three-dimensional domain swept
by rotating that cross-section about its vertical axis. The external
boundaries (in green, labeled bath \& guard electrode) are the
(voltage-clamp) electrode surfaces kept at controlled electrical
potentials. The blue zones represent aqueous baths (labeled with a
dielectric coefficient $\epsilon_{\mathrm{w}}=80$). The pink zone is a
region of small dielectric coefficient (labeled with
$\epsilon_{\mathrm{m}}=2$) that represents the lipid membrane. The
brown zone (labeled S1--S3 \& S4) represents the region of the
channel protein that we model; this region is assigned a dielectric
coefficient of $\epsilon_{\mathrm{p}}=4$ unless otherwise noted. These
dielectrics are piecewise uniform and therefore have sharp boundaries
(solid black lines). Point source charges representing protein charges
of interest are embedded in the region of protein
dielectric. Variation of their placement is part of this study and
will be detailed later. The protein region as seen here represents the
matrix of the S4 helix as a central cylinder, surrounded by the other
parts of the channel that also create a dielectric environment
different from the dielectric environment of the membrane
lipid. Included in the protein region are the invaginations which
allow the baths to extend into the planes defining the lipid phase of
the membrane. The radius of the S4 dielectric domain is $1$~nm
(\textalpha helix) or $0.98$~nm ($3_{10}$ helix).

\begin{figure*}
  \mbox{
    \includegraphics[width=0.5\linewidth]{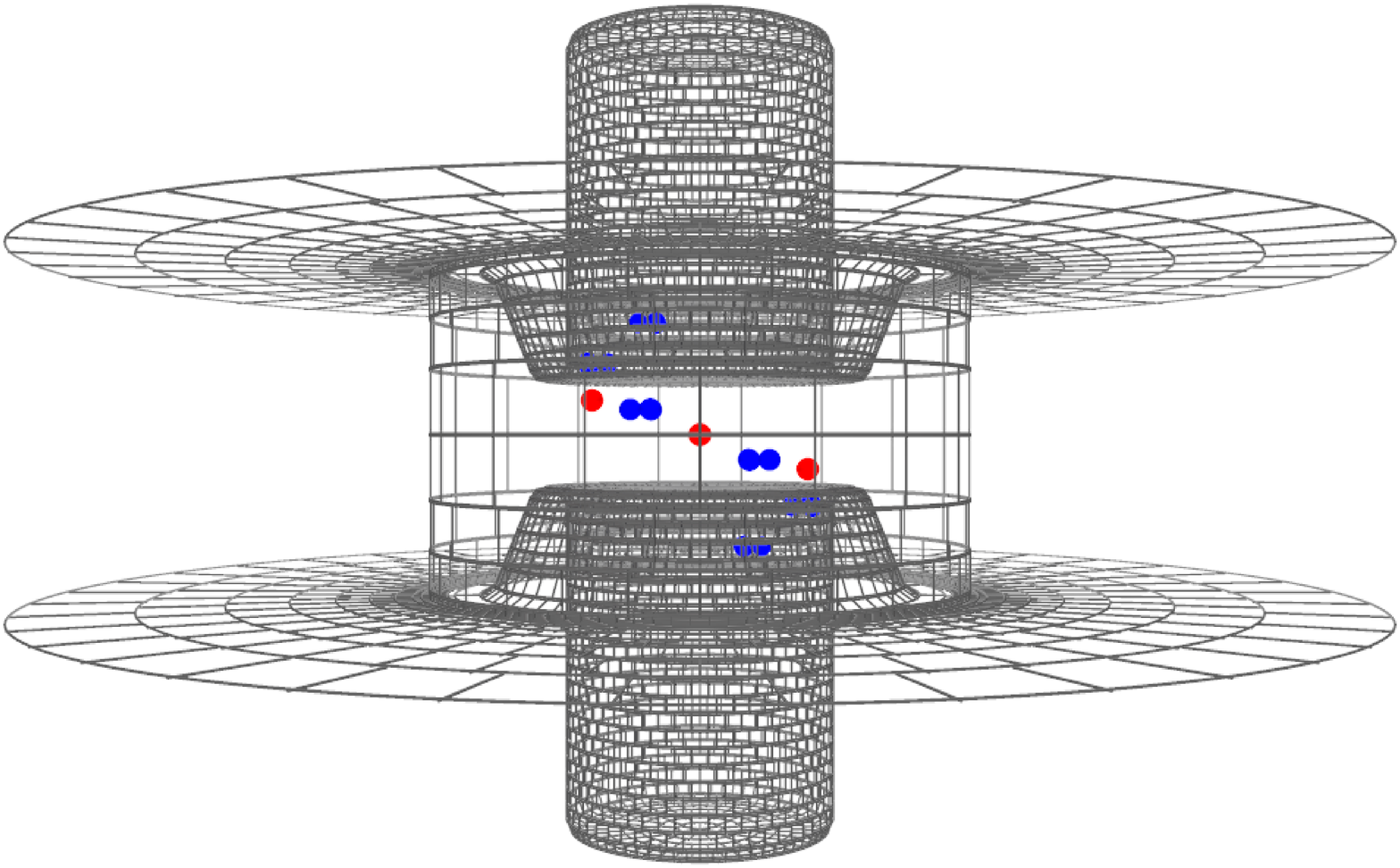}
    \includegraphics[width=0.5\linewidth]{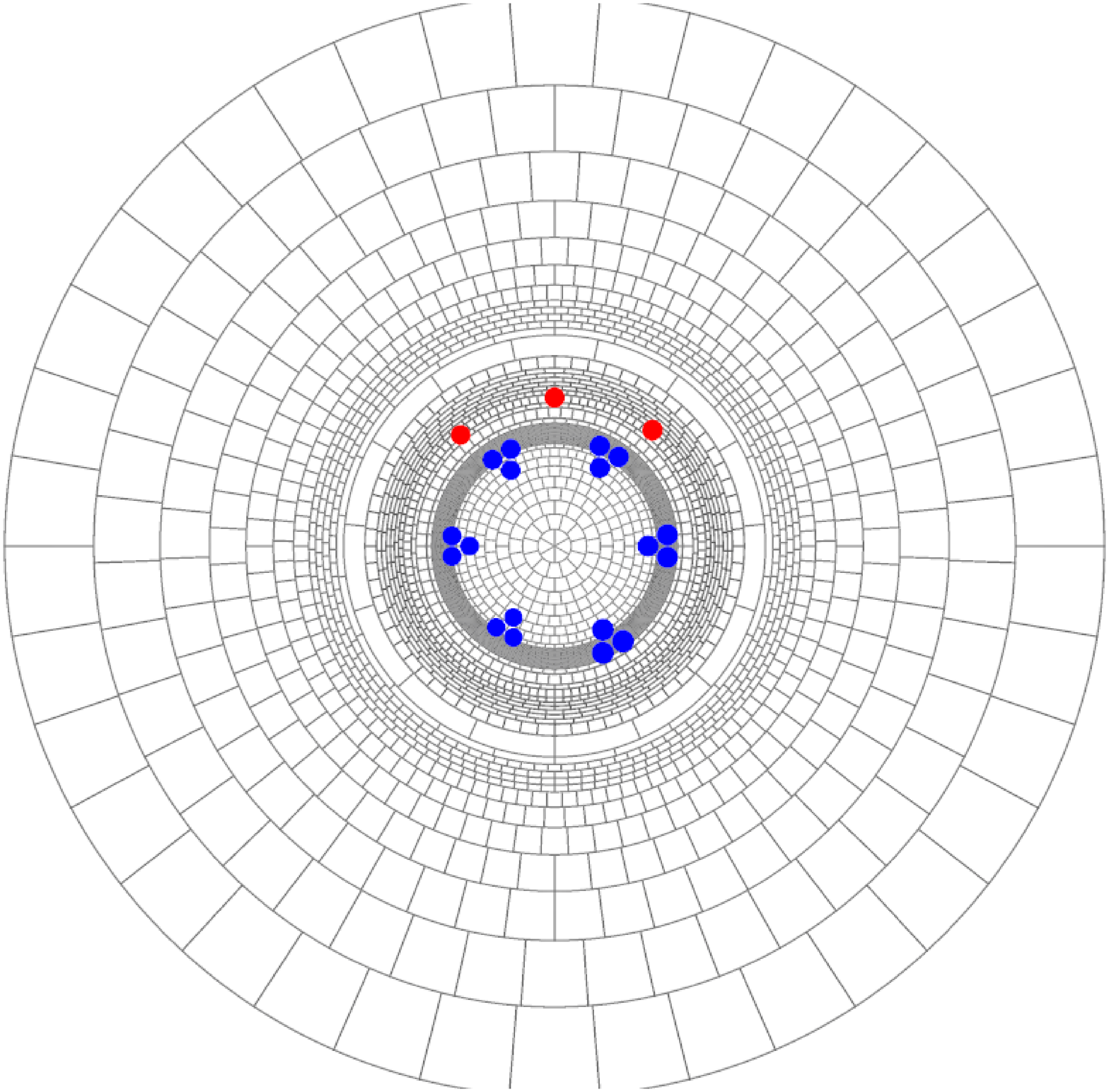}
  }
  \caption{\emph{VS charge positions and dielectric boundary
      surfaces in the \emph{\textalpha}-helical model.}  \emph{Blue
      (dark gray) symbols}: S4 charges, each represented as three
    point source charges of $+(1/3)e_0$; \emph{Red (light
      gray) symbols}: countercharges in the S2 and S3 segments, each
    represented as a single point source charge of
    $-1e_0$. The dielectric boundary surface is divided into
    curved tiles whose magnitudes are varied depending on their
    distance from the point source charges and local curvature.  This is
    the surface grid used in solving the induced-charge calculation.
    The figure is drawn using perspective-enhancing features. See also
    Fig.~\ref{fig:movie:standard}.}
  \label{fig:helix}
\end{figure*}

Point source charges representing protein charges are arranged at a
minimal distance of $0.2$~nm from the protein/water boundary. The
charged guanido group of each arginine residue of the S4 segment is
represented as three point source charges of $+(1/3)e_0$ on a
circle of radius $0.122$~nm [blue (dark gray) spheres in
Fig.~\ref{fig:helix}]. The centers of the S4 arginine charges are
arranged on a helix defined by arginine side chains on an \textalpha-
or $3_{10}$ helix backbone, where every third amino acid is an
arginine. For the \textalpha helix, charged residues are separated by
$0.45$~nm in the transmembrane direction and 60\textdegree{} leftward
around the helix; for the $3_{10}$ helix, charged residues are
separated by $0.6$~nm and $0$\textdegree.

Dimensions of the simulation surfaces are varied within those
constraints as defined in Figs.~\ref{fig:cell}(B)
and \ref{fig:cell}(C).  Figure~\ref{fig:cell}(B) maps the topology of
the protein and membrane surface of Fig.~\ref{fig:cell}(A) onto
metrics for simulations, defined in Fig.~\ref{fig:cell}(C). Since the
surfaces in the system are radially symmetrical and smooth, the system
is defined by a set of inflection points with their curvature on the
left half of the system to be simulated. Since the gating canal is
symmetrical in these models, this further reduces to the upper half of
the left side.

We analyze two degrees of freedom for the movement of the S4 segment:
the curve on which (triplets of) S4 charge centers are aligned can be
both translated along the helix axis and rotated about that axis. The
model S4 charges thus move like parts of a solid body. Negatively
charged residues contributed by the S2 and S3 transmembrane segments
of the natural VS are modeled as point source charges of
$-1$~$\text{e}_0$ fixed on a curve parallel to the curve on which S4
charges are centered [red (light gray) spheres in
Fig.~\ref{fig:helix}]; the offset from the helix axis of the
countercharge curve is $0.466$~nm larger than the radius of the curve
of the centers of the (triplets of) S4 charges [see
Fig.~\ref{fig:cell}(C)]. The axial and angular intervals between
countercharges are discussed in
Sec.~\ref{sec:results}. countercharges are stationary in their
assigned positions.

The electrodes encapsulating the simulation cell serve three purposes:
\begin{enumerate}
\item The bath electrodes provide Dirichlet boundary conditions
  corresponding to a voltage clamp. 
\item The bath electrodes substitute for screening by bath ions of
  uncompensated protein charge. Screening by the ions in an aqueous
  bath is equivalent to the screening provided by charge on a metal
  foil placed in the water a distance away from the protein
  boundary. In the Debye-Hückel theory, an electrode distance of
  \textsim$0.8$~nm corresponds to the physiological bath ionic strength;
  thus the electrode location shown in Fig.~\ref{fig:cell}(A)
  corresponds to a bath solution in the low millimolar range. An
  alternate configuration, a simulation cell with the bath regions
  omitted and the electrodes placed directly on the membrane and
  protein boundaries, would establish screening at the Onsager limit
  approached at exceedingly large ionic strength. In this way, the
  possible range of screening effects can in principle be determined
  without including explicit bath ions in the simulation.
\item At the surface where the membrane region meets the cell
  boundary, a set of guard electrodes forming rings around the cell
  maintain a graded far potential varying, from ring to ring, between the
  potentials applied at the inner and outer bath electrodes. These
  guard electrodes impose at the membrane edge a far potential similar
  to that existing in a macroscopic system.
\end{enumerate}

Translational and rotational motions of the S4 helix are simulated by
allowing the ensemble of S4 charges to slide within the dielectric
domain of the protein shown in Fig.~\ref{fig:cell}. The protein
dielectric itself does not slide with the charges. With the protein
dielectric extended far enough into the baths, keeping the S4
dielectric stationary has negligible electrostatic consequences
because the dielectric of the model is uniform within the protein
domain. Simulating S4 motion in this way reduces the computational effort
of model exploration by several orders of magnitude.

\section{Methods}

\subsection{Electrostatics}

We are concerned with the electrostatic interactions among charged
groups of the VS protein, electrode charges and charges induced on
sharp dielectric boundaries. In solving the electrostatics, we take
advantage of the fact that all other charges besides the point source
charges of the protein are distributed on a few boundary surfaces
rather than distributed throughout a volume. The primary task of
calculating electrostatic interactions consists in determining the
charge distributions on the electrode and dielectric boundaries,
distributions which are initially unknown for a given configuration of
protein charges and applied voltage.

\subsubsection{Computation of unknown charges}

\Citet{boda:2006} have described and tested a boundary element method
(the induced charge calculation) for computing the charge distribution
on the dielectric boundaries of a system consisting of point source
charges and linear isotropic dielectrics with sharp boundaries. We
include an additional electrostatic element, the `electrode': an
infinitesimally thin conductive foil charged to a prescribed electric
potential using an external source. In our system, this surface is
also a dielectric boundary between the simulation cell and the
dielectric surrounding the cell (e.g., vacuum). Thus the charge of an
electrode is in part induced charge (as in the paper
by \citeauthor{boda:2006}) and source charge (to fix the prescribed
potential). In the following, we show how the total electrode charge
can be calculated without needing to separately calculate its
parts. The electrode charge calculation complements the induced charge
calculation, making it possible to calculate all the initially unknown
charges in the system.

The spatial density of source charge $\rho^{\text{src}}$ in our system
is composed of the point source charges $q_k^{\text{src}}$ of the VS
protein located at positions $\mathbf{r}_k$ and source charge
distributed on the electrodes $\mathcal{E}$ with surface density
$\sigma^{\text{src}}(\mathbf{r}\in\mathcal{E})$.

The polarization charge density induced at any location $\mathbf{r}$
in a dielectric is
\begin{equation}
\rho^{\text{ind}}(\mathbf{r}) =
  \frac{1 - \epsilon(\mathbf{r})}{\epsilon(\mathbf{r})}
      \rho^{\text{src}}(\mathbf{r})
   -  \epsilon_0
      \frac{\nabla \epsilon(\mathbf{r})}
       {\epsilon(\mathbf{r})}
      \cdot \mathbf{E}(\mathbf{r}),
\label{eq:indcharge}
\end{equation}
where $\epsilon$ is the generally location-dependent dielectric
coefficient, $\epsilon_0$ the permittivity of the vacuum, and
$\mathbf{E}$ the electric field strength produced by all
(source and induced) charges in the system. This relation follows from
Poisson's equation (including polarization) and the constitutive
relation describing polarization in a linear, isotropic dielectric
\citep{boda:2006}. The first term on the right-hand side of
Eq.~(\ref{eq:indcharge}) describes the charge induced on the surfaces of
the volume element at $\mathbf{r}$ if the element contains source
charge. The second term describes charge induced in the volume element
by the electric field if the dielectric coefficient at $\mathbf{r}$
has a non-zero gradient.

For our charge calculations, we combine collocated source and induced
charges into an effective charge for computing the field and
potential. The effective charge density associated with a known source
charge density $\rho^{\text{src}}(\mathbf{r})$ embedded in
a dielectric [described by a locally uniform
$\epsilon(\mathbf{r})$] is
\begin{equation}
\rho^{\text{eff}}(\mathbf{r}) = \frac{\rho^{\text{src}}(\mathbf{r})}{\epsilon(\mathbf{r})}.
\label{eq:effcharge}
\end{equation}
The effective charge density of an electrode includes both
contributions to the induced charge described by
Eq.~(\ref{eq:indcharge}), as well as the source charge.

The dielectric boundaries $\mathcal{B}$ inside the simulation cell
(marked in heavy \emph{black} in Fig.~\ref{fig:cell}; all lines but
the electrodes) do not carry source charge. However, the electric field in
the simulation cell induces the charge density
$\sigma^{\text{ind}}(\mathbf{r})$ at locations $\mathbf{r} \in
\mathcal{B}$. This induced charge density is initially unknown.

The field strength $\mathbf{E}$ and potential $V$ in our system are
produced by the superposition of the fields and potentials of the
source and induced charges:
\begin{equation}
\begin{split}
4\pi\epsilon_0\;\mathbf{E}(\mathbf{r}) & =
   \sum_k q_k^{\text{eff}}
     \frac{\mathbf{r}-\mathbf{r}_k}{|\mathbf{r}-\mathbf{r}_k|^3} \\
 & + \int_{\mathcal{B}}\!\sigma^{\text{ind}}(\mathbf{r}')
     \frac{\mathbf{r}-\mathbf{r}'}{|\mathbf{r}-\mathbf{r}'|^3}
     \;da' \\
 & + \int_{\mathcal{E}}\!\sigma^{\text{eff}}(\mathbf{r}')
     \frac{\mathbf{r}-\mathbf{r}'}{|\mathbf{r}-\mathbf{r}'|^3}
     \;da',
\end{split}
\label{eq:field}
\end{equation}

\begin{equation}
\begin{split}
4\pi\epsilon_0\;V(\mathbf{r}) & =
   \sum_k q_k^{\text{eff}}
     \frac{1}{|\mathbf{r}-\mathbf{r}_k|} \\
 & + \int_{\mathcal{B}}\!\sigma^{\text{ind}}(\mathbf{r}')
     \frac{1}{|\mathbf{r}-\mathbf{r}'|}\;da' \\
 & + \int_{\mathcal{E}}\!\sigma^{\text{eff}}(\mathbf{r}')
     \frac{1}{|\mathbf{r}-\mathbf{r}'|}\;da',
\end{split}
\label{eq:potential}
\end{equation}
where $da'$ is the area of the surface element at location $\mathbf{r}'$.

The unknown induced surface charge density on the dielectric boundary
is related to the field strength by \citep{boda:2006}
\begin{equation}
  \sigma^{\text{ind}}(\mathbf{r}) = 
  - \frac{\Delta \epsilon (\mathbf{r})}
         {\bar{\epsilon}(\mathbf{r})}
  \,\epsilon_0
  \,\mathbf{n}(\mathbf{r}) \cdot
  \mathbf{E}(\mathbf{r}),
\label{eq:chargeB}
\end{equation}
where $\mathbf{r}$ is any location on the surface $\mathcal{B}$,
$\Delta \epsilon(\mathbf{r})$ is the difference in the dielectric
coefficient across the dielectric boundary in the normal direction
$\mathbf{n}(\mathbf{r})$, and $\bar{\epsilon}(\mathbf{r})$ is the mean
of the dielectric coefficients at the boundary location.

The potential at any location $\mathbf{r}$ on the electrode surfaces
$\mathcal{E}$ has a value $V^{\mathrm{VC}}(\mathbf{r})$
imposed by the voltage clamp:
\begin{equation}
V(\mathbf{r}) = V^{\mathrm{VC}}(\mathbf{r}).
\label{eq:chargeE}
\end{equation}
Inserting the expression for the electric field strength from
Eq.~(\ref{eq:field}) into Eq.~(\ref{eq:chargeB}) and inserting the
expression for the electric potential from Eq.~(\ref{eq:potential}) into
Eq.~(\ref{eq:chargeE}) yields two integral equations in terms of both
$\sigma^{\text{ind}}(\mathbf{r}\in\mathcal{B})$ and
$\sigma^{\text{eff}}(\mathbf{r}\in\mathcal{E})$. The initially unknown
charge densities on the dielectric and electrode boundaries are the
joint solution of these two integral equations.

To solve the integral equations, we follow the method of
\citet{boda:2006}. The surfaces $\mathcal{B}$ and $\mathcal{E}$ are
subdivided into curved surface elements. The unknown charge densities
are approximated as uniform on each surface element. The two integral
equations then become one system of linear equations in terms of the
unknown charge densities of a finite number of surface elements.

The inhomogeneity of surface charge in our simulation is greatest
where point source charges of the VS protein are close to a dielectric
boundary or electrode. For computational efficiency, we vary the size
of the surface subdivisions depending on the distance from the point source
charges. A typical surface grid (comprising \textsim7000 surface
elements) is shown in Fig.~\ref{fig:helix} for the dielectric surfaces
in the simulation cell. The electrode surfaces (not included in
Fig.~\ref{fig:helix}) are subdivided into \textsim1700 relatively
large elements because of their distance from point source charges.

The computation of the unknown charges on the surface elements
involves solving a linear equation system in terms of as many unknowns
$(N)$ as there are surface elements. The coefficient matrix of this
system is dense, therefore the LU-decomposition time increases with
$\mathcal{O}(N^3)$. This computational disadvantage is greatly
alleviated by the fact that $LU$-decomposition of the coefficient
matrix needs to be done only once for a given combination of surface
geometry and dielectric coefficients. When the S4 charges are moved,
solutions to this system of equations are obtained by
back-substitution using the same LU-decomposed matrix. That
$\mathcal{O}(N^2)$ operation (back-substitution) is required for each
sampled configuration of VS charges but not for each applied voltage
tested, as described later.

\paragraph*{Accuracy of charge calculation. } The divergence theorem
(Gauss's theorem) states that
\begin{equation}
\oint_{\mathcal{S}}\!\epsilon(\mathbf{r})\,\epsilon_0
   \mathbf{E}(\mathbf{r}) \cdot 
   \mathbf{n}(\mathbf{r})\;da =
   \int_{\mathcal{V}}\!\rho^{\text{src}}(\mathbf{r})\;d\tau
   \label{eq:gauss}
\end{equation}
where $\mathcal{S}$ is the closed surface around the volume
$\mathcal{V}$, $da$ is the area of the surface element located at
$\mathbf{r} \in \mathcal{S}$, $\mathbf{n}(\mathbf{r})$ is the normal
unit vector for that surface element, $d\tau$ is the volume of the
space element located at $\mathbf{r} \in \mathcal{V}$, and
$\rho^{\mathrm{src}}$ is the source charge density inside the volume.

Gauss's theorem provides a sum-rule test for our solution of the
electrostatics that is applicable to the specific geometry used in a
simulation. We verify the theorem at the dielectric boundary of the VS
protein because: 1) that boundary is closest to the protein source
charges of interest; and 2) the field at that boundary produces the
largest density of induced charge in our system. The volume integral
in Eq.~(\ref{eq:gauss}) yields the known algebraic sum of the point
source charges assigned to the VS residues. The surface integral can
be expressed in terms of the charges induced by the normal field at
the discretized dielectric boundary. We consider the residue given by
the difference of the surface and volume integrals as a measure of our
numerical error:

\begin{equation}
  Q_{\text{error}}  = 
  -\sum_j \frac{\epsilon_{\mathrm{p}} \epsilon_j}{\epsilon_j - \epsilon_{\mathrm{p}}}
          \sigma^{\text{ind}} a_j - \sum_k q_k^{\text{src}},
\label{eq:gausserror}
\end{equation}
where $a_j$ is the area of the protein surface element $j$, $\epsilon_j$
the dielectric coefficient on the out-facing side of $a_j$, and
$\sigma^{\text{ind}}$ the induced charge density computed to solve the
electrostatics.

\begin{figure}
  \includegraphics{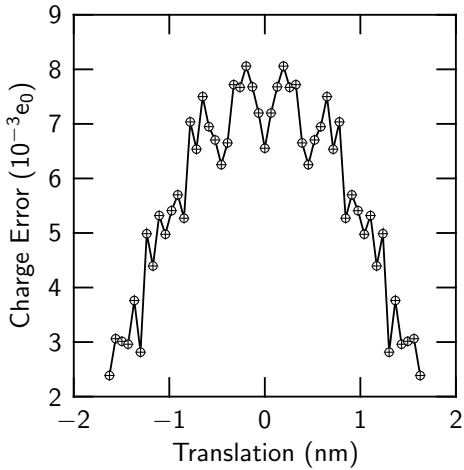}
  \caption{\emph{Test of numerical accuracy using Gauss's theorem.}
    $Q_{\mathrm{error}}$ as defined by Eq.~(\ref{eq:gausserror}) is
    plotted versus S4 translation [Eq.~(\ref{eq:randvarrot}) is applied
    to calculate the expectation value over the rotational degree of
    freedom].  Normal electric flux is integrated over the surface of
    the protein (brown region in Fig.~\ref{fig:cell}, labeled
    S1--S4). The protein region contains a net charge of
    $+3e_0$ (six positive S4 charges and three negative
    countercharges).  The charge error is computed for three applied
    voltages ($-100$~mV, circles; $0$~mV, line; $100$~mV, crosses).}
  \label{fig:gausstest}
\end{figure}

The error in the induced charge calculation is likely to vary as S4
charges are moved in a simulation since the distances between the S4
charges and the dielectric boundary vary. Figure~\ref{fig:gausstest}
shows the error in induced charge calculation for the full range of S4
translational positions sampled in a typical simulation. The error is
\textleq$0.008$~$\text{e}_0$ of the actual net charge of
$3$~$\text{e}_0$ assigned to the VS in this simulation. The results of
our charge calculation are thus in good agreement with Gauss's
theorem, as they must be. This test was performed for every
simulation.

\subsubsection{Computation of charge displacement and electrostatic energy}

When the charges of the VS change position, the electric flux toward
one bath electrode generally increases by the same amount as the
electric flux decreases toward the other bath electrode. To maintain a
constant voltage between the two electrodes, charge has to be moved
externally between the electrodes. This charge is the experimentally
measured displaced gating charge. In principle, one can measure
displaced charge in a simulation by monitoring electric flux across a
surface surrounding a bath electrode. A more efficient method is
provided by the Ramo-Shockley (RS) theorem
\citep{shockley:1938,ramo:1939}; for an application to ion channels
see \citep{nonner:2004:rs}). The RS theorem lays the groundwork as
well for an efficient method of computing the electrostatic energy of
VS configurations when the applied voltage is varied
\citep{he:2001}. The RS theorem is applicable to systems containing
linear dielectrics.

The RS theorem can be formulated for the configuration of electrode
potentials in this study. We apply equal and opposite potentials
$V_{\mathrm{m}}/2$ and $-V_{\mathrm{m}}/2$ to the internal and
external bath electrodes to create a membrane voltage $V_{\mathrm{m}}$
(defined as internal minus external potential). We determine the
displaced charge in a simulation in two steps:
\begin{enumerate}
\item Set all point source charges to zero and apply $+1/2$ V at
  the internal and $-1/2$ V at the external bath electrode. Solve
  for the unknown electrode and induced boundary charges. From these
  charges, an electric potential $V_0(\mathbf{r})$ can be computed for
  any geometrical location $\mathbf{r}$ of the simulation cell.
\item In a simulation with the actual point source charges $q_k$
  present and arbitrary potentials $+V_{\mathrm{m}}/2$ and
  $-V_{\mathrm{m}}/2$ applied at the internal and external bath
  electrodes, determine the displaced charge $Q_k$ from the relation
\begin{equation}
Q_k = -q_k\;V_0(\mathbf{r}_k)/(1\text{ V})
\label{eq:rs}
\end{equation}
Note that $Q_k=0$ for all geometrical positions $\mathbf{r}_k$ where
$V_0(\mathbf{r}_k)=0$, and $Q_k$ varies between $-q_k/2$
and $+q_k/2$ as the position of $q_k$ is varied from the internal to
the external bath electrode.

When several point source charges are in the simulation, the total
displaced charge is the algebraic sum of the displaced charges defined
by Eq.~(\ref{eq:rs}) for each point source charge:
\begin{equation}
Q = \sum_k Q_k
\label{eq:rsall}
\end{equation}

\end{enumerate}
Step 1 is executed once for the chosen configuration of electrodes and
dielectrics in the simulation (including dielectric
coefficients). Step 2 is executed once for each configuration of point
source charges. Since the displaced charge determined in step 2 is
invariant with respect to the potentials applied at the electrodes, it
needs to be calculated only once for each point-charge configuration,
regardless of variation in the electrode potentials.

The RS theorem also makes it possible to compute the electrostatic
energy of a VS configuration with the algebraic sum of two terms:
\begin{equation}
W = W_1 + W_2,
\label{eq:ework}
\end{equation}
determined by separate calculations:
\begin{enumerate}
\item In a simulation which includes the point source charges $q_k$ at
  positions $\mathbf{r}_k$, impose the potential $V_{\mathcal{E}}=0$
  on all electrodes and compute the self-energy
\begin{equation}
W_1 = \frac{1}{2} \sum_k q_k V^{V_{\mathcal{E}}=0} (\mathbf{r}_k) 
\label{eq:nullenergy}
\end{equation}
where $V^{V_{\mathcal{E}}=0} (\mathbf{r}_k)$ is determined by
Eq.~(\ref{eq:potential}), excluding self-interactions for each source
charge $q_k$.

\item Calculate the displaced charge $Q$ corresponding to the point source
  charges and their positions using Eq.~(\ref{eq:rsall}). For the
  imposed voltage $V_{\mathrm{m}}$, calculate:
\begin{equation}
W_2 = -QV_{\mathrm{m}}
\label{eq:pickedup}
\end{equation}

\end{enumerate}
Step 1 of this procedure is executed once for each sampled
configuration of point source charges. Step 2 is executed repeatedly
for each applied voltage that is tested.

The calculations of displaced charge and electrostatic energy via
Eqs.~(\ref{eq:rsall}) and (\ref{eq:ework}) are independently verified by
computing the electrostatic energy by the path integral of the
electric force acting on the charges $q_k$ of the VS as those charges
move from $\mathbf{r}_k'$ to $\mathbf{r}_k'$:
\begin{equation}
  \label{eq:ework2}
  \Delta W = \sum_k q_k
    \int_{\mathbf{r}_k'}^{\mathbf{r}_k'}
    \hskip-1.5ex\mathbf{E}(\mathbf{r}_k)
    \cdot d\mathbf{r}_k
\end{equation}
Here, the electric field is the field of all charges in the system except the
charge $q_k$ itself, as defined by Eq.~(\ref{eq:field}).

\begin{figure}
  \includegraphics{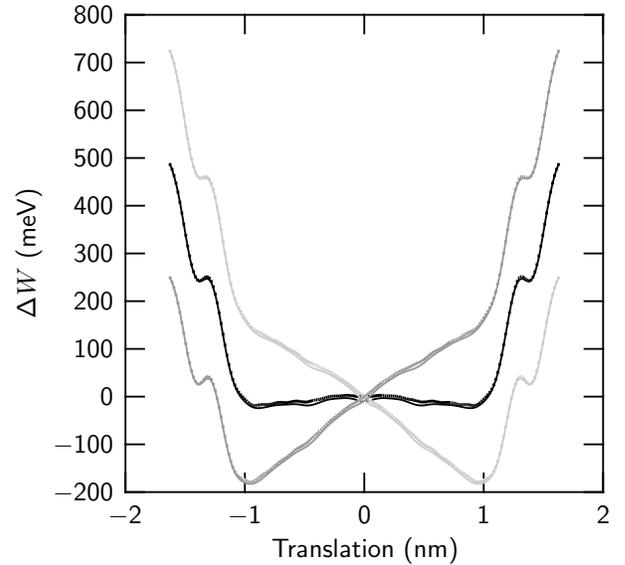}
  \caption{\emph{Electrostatic energies computed via different
      routes.} Energy scanned for a fixed diagonal path through the
    two dimensions (translation as shown on abscissa, rotation from
    $-180$\textdegree to $+180$\textdegree). Results of Eq.~(\ref{eq:ework})
    (\emph{dots}) and Eq.~(\ref{eq:ework2}) (\emph{Lines}, aligned to
    dots at the negative end of the path). Applied voltage
    $V_{\mathrm{m}}$: $-100$~mV (dark gray), $0$~mV (black);
    $+100$~mV (light gray).}
\label{fig:Ebypath}
\end{figure}

Figure~\ref{fig:Ebypath} shows this control for three different applied
voltages over a prescribed (diagonal) path through the translational
and rotational dimensions of the range of S4 motion typically sampled
by us.  There is good agreement between the energies computed using
the Ramo-Shockley theorem (dots) or through the path integral of force
(lines).

\subsection{Statistical mechanics}\label{sec:stats}

Displaced gating charge is experimentally measured from ensembles of
channels and thus is an ensemble average. Our electrostatic
calculations yield both the displaced charge and the electrostatic
part of the configurational energy for a given configuration of a
simulated VS model. We consider whole-body movements of S4 charge in
two degrees of freedom: translation along the S4 axis and rotation
about that axis. Our computational method is efficient enough to allow
systematic sampling of this configuration space. We represent each
dimension by 51 equally spaced grid nodes and compute the electrostatic
energy for the 2601 nodes of the two-dimensional space.

The energy samples define a canonical partition function describing
the consequences of the electrostatics on the distribution of an
ensemble in the discretized configuration space:
\begin{equation}
\mathcal{Q} = \sum_{i,j} e^{-\Delta W_{ij}/k_BT},
\label{eq:partitionfct}
\end{equation}
where $i$ and $j$ are the indices of the rotational and translational
discrete positions; $\Delta W_{ij}$ is the electrostatic configurational
energy of the voltage sensor at translational position $i$ and
rotational position $j$; $k_B$ is the Boltzmann constant; and $T =
298.15$~K is the absolute temperature. The sampled rotational range is
$360$\textdegree, and a typical translational range is $-1.625$~nm to
$+1.625$~nm relative to the central position of the S4 charges in
\textalpha-helical models ($\pm2.102$~nm for $3_{10}$-helical
models). Restricting the configuration space this way is equivalent to
including hard-wall potentials into the Hamiltonian [Eq.~(\ref{eq:ework})].

The probability of a VS configuration for a particular translation $i$
and rotation $j$ is
\begin{equation}
  \label{eq:dist}
  P_{ij} = \frac{1}{\mathcal{Q}}e^{-\Delta W_{ij}/k_BT}
\end{equation}
and the expectation (mean) value of a random variable $X$ is
\begin{equation}
  \begin{split}
    \langle X\rangle &= \sum_{i,j} X_{ij}P_{ij} \\
    &= \frac{1}{\mathcal{Q}}\sum_{i,j} X_{ij} e^{-\Delta W_{ij}/k_BT}.
  \end{split}
\label{eq:randvar}
\end{equation}
We also determine the expectations of random variables over the rotational
degree of freedom for a particular translational position $i$:
\begin{equation}
  \langle X_i\rangle = \frac{\sum_{j} X_{ij} e^{-\Delta W_{ij}/k_BT}}{\sum_je^{-\Delta W_{ij}/k_BT}}
  \label{eq:randvarrot}
\end{equation}

The goal of our simulations is to study the energetics of the movement
of the S4 segment for two degrees of freedom. We are concerned
only with variation of the Hamiltonian due to changes in S4 position for
this space. Since these simulations are done for varied settings of an
external parameter (applied voltage), the energy at the reference
position for $\Delta W$ must be invariant with respect to applied
voltage.

In our simulations, a voltage $V_{\mathrm{m}}$ is applied across the
bath electrodes by imposing the potentials $+V_{\mathrm{m}}/2$ and
$-V_{\mathrm{m}}/2$ at the internal and external bath
electrodes. Hence, there are locations in the simulation cell where
the potential $V$ due to the applied field and the displaced charge
$Q_k$ for any point charge there are zero for any $V_{\mathrm{m}}$
[Eq.~(\ref{eq:rs})]. Since we are concerned with the position of the S4
segment which bears point charges $q_k$ at several locations, we
define the S4 reference position such that the total displaced charge
there is zero. By Eq.~(\ref{eq:rsall}) and due to the polarity of the
fields applied, there are positions for the S4 segment where $Q$ is
zero, and therefore the energy for those configurations is invariant
with regard to $V_{\mathrm{m}}$. Because of the symmetries in this
study, these reference positions coincide with the origin of the
translational axis ($z=0$~nm for any rotation $\phi$; we choose
$\phi=0$).

\subsection{Online Supplemental Materials}

Figures~\ref{fig:movie:first} \& \ref{fig:movie:last} and the associated
Supplemental Animations~1~\cite{peyser:2012:supp:1} \&
2~\cite{peyser:2012:supp:2} illustrate VS geometry and movement for
the simulations presented in this paper. They show the expectation(s)
of position for VS charges superimposed over the distribution of
charge density. In the animation, voltage is changed in a ramp from
$-100$ to $+100$~mV. Note that these animations are of the VS without
an external load attached --- as in Sec.~\ref{sec:idle} rather
than Sec.~\ref{sec:working}.

\section{Results and Discussion}\label{sec:results}

\begin{figure}
  \includegraphics{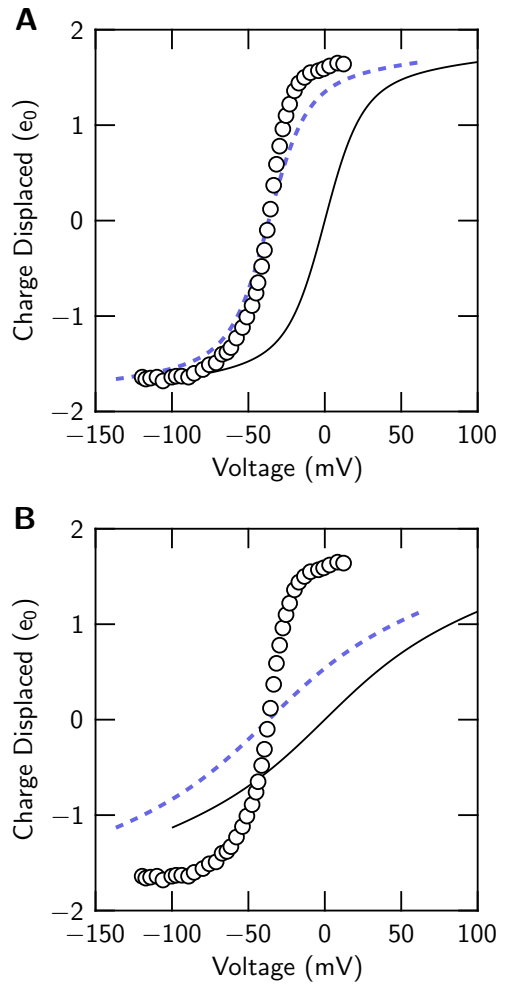}
  \caption{\emph{Simulated charge-voltage relations
      (solid lines) compared to experiment (symbols).} (A)
    $\alpha$-helical S4 segment; (B) $3_{10}$-helical S4 segment. The
    experimental relation applies to \emph{Shaker} $\text{K}^+$
    channels \citep{seoh:1996}.  \emph{Dashed lines:} computed curves
    shifted to match the midpoint of the experimental curve (shift is
    $1/2$ of the experimental range).}
  \label{fig:qvsbothhelices}
\end{figure}

The relationship between the expectation of displaced charge and the
applied membrane voltage in a sliding-helix model of an individual VS
domain is shown in Fig.~\ref{fig:qvsbothhelices}(A). The solid line
represents the computed relation. Open circles represent the
relationship experimentally observed in \emph{ShakerB} K$^+$ channels
\citep[Fig.~2A in][]{seoh:1996} --- the experimental charge per
channel was divided by the number of channel monomers (4, where each
includes one VS domain).  Three observations can be made by comparing
the two relations: (1) the total amounts of charge that can be moved
by large changes of voltage are similar, \textsim$3$ elementary
charges per VS domain; (2) the slopes of the two relations are
similar; and (3) a shift along the voltage axis is needed to align the
midpoint of the computed relation with the midpoint of the experimental
relation (dashed line).

The VS charges in the simulation for Fig.~\ref{fig:qvsbothhelices}(A)
are arranged according to an \textalpha-helical geometry
(Figs.~\ref{fig:helix} \& \ref{fig:movie:standard}).  An analogous
simulation in which the VS charges conform to a $3_{10}$-helical
geometry (Fig.~\ref{fig:movie:310}) yields a different result
[Fig.~\ref{fig:qvsbothhelices}(B)]. With regard to the experimental
charge-voltage relation, the $3_{10}$ model yields less total charge
movement over the tested voltage range and a smaller maximal slope.

The models giving the charge-voltage relations of
Fig.~\ref{fig:qvsbothhelices} involve idealized domain geometries and
use the dielectric coefficient $\epsilon_{\mathrm{p}}=4$. These are
initial choices made in this study --- no variations were made to
produce more realistic predictions. The geometries of S4 charges were
idealized to conform with two kinds of helices. The countercharges
($3$) were arranged on a spiral (\textalpha helix) or straight line
($3_{10}$ helix) paralleling the curve of the S4 charges. The only
parameter optimized in light of the experimental results was the
countercharge spacing (the translational and angular intervals between
countercharges). In the models used in this paper, countercharges
are spaced at $2/3$ the intervals of the S4 charges, so that at most
one S4 charge lines up with a countercharge for any particular S4
position.

The simulations reported in Fig.~\ref{fig:qvsbothhelices} indicate
that voltage sensing by a sliding-helix is robust from an engineering
point of view. In these generic models, either the \textalpha- or
$3_{10}$-helical S4 structure can produce voltage-dependent charge
displacement, even though the two structures involve distinct
configurations of the protein charges. The two forms of helix generate
different responses to voltage in the tested models, but neither form
produces catastrophic failure. The biological channels whose
structure-function relation we seek to ultimately understand in
engineering terms themselves show robustness: different channel types
exhibit a wide range of response to voltage while remaining functional
under many mutations that cause their response to change, and yet they
all use a common architecture.  On these grounds, useful insights are
expected from analyzing a mesoscale physical model.

\subsection{Energetics of `idling' voltage sensors}\label{sec:idle}

The mobile charges of the VS model lie within the electric field of
the charges on the electrodes, the stationary countercharges, and the
charges induced in the dielectrics. We model those mobile charges as
parts of a solid body with two degrees of freedom of solid-body
motion: translation along the S4 axis and rotation about that
axis. Energy is computed on a grid over this configuration space
[Eq.~(\ref{eq:ework})]. From the electrostatic energy map, a partition
function is constructed [Eq.~(\ref{eq:partitionfct})], and from the
partition function, statistics of random variables
[Eq.~(\ref{eq:randvar})] are predicted. The energy map thus defines the
expectations of observables such as the gating charge displacement
(Fig.~\ref{fig:qvsbothhelices}). This energy map is computed for each
tested value of applied voltage. These maps and the resulting partition
functions are outputs of the model system.

\begin{figure*}
  \hspace*{3em}\includegraphics{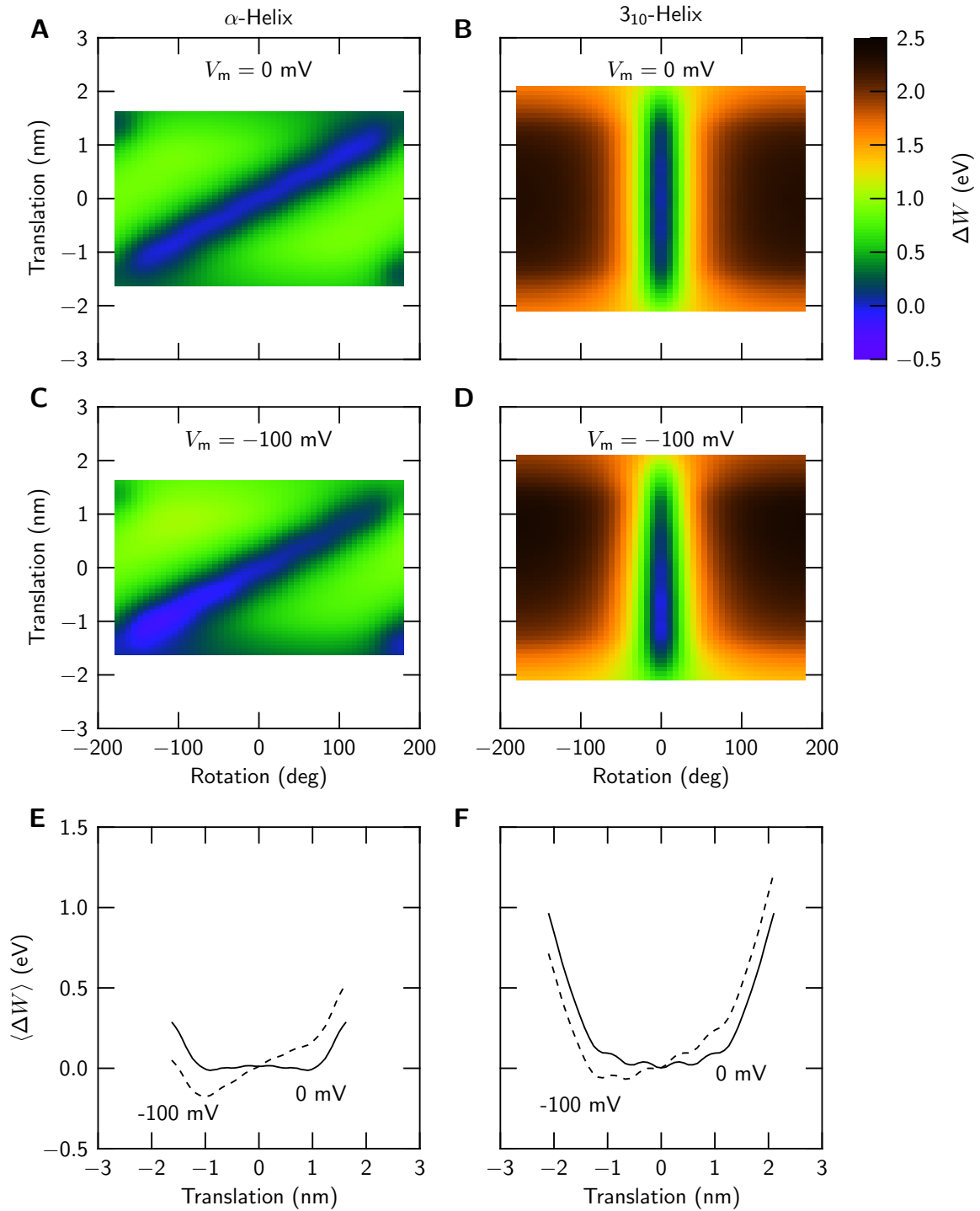}
  \caption{\emph{Energetics of voltage sensing.} (A, C, E)
    \textalpha-helical S4 segment; (B, D, F) $3_{10}$-helical S4
    segment. Pseudo-color maps: electrostatic energy for applied
    voltages $0$~mV (A, B) and $-100$~mV (C, D). (E, F): Expectation
    of energy over the rotational degree of freedom versus
    translation. Applied voltage $V_{\mathrm{m}}=0$~mV (solid line);
    $V_{\mathrm{m}}=-100$~mV (dashed line).}
  \label{fig:Ebothhelices}
\end{figure*}

Pseudo-color maps of electrostatic energy in the two dimensions
computed for applied voltages of $0$~mV and $-100$~mV are shown in
Fig.~\ref{fig:Ebothhelices}. The translational and rotational origins
correspond to the central S4 position shown in Fig.~\ref{fig:helix}
for the \textalpha helix (change in energy is relative to those
origins, where $Q$ is zero). Because of the symmetry of these models,
the map for $+100$~mV (not shown) is a mirror image of the map for
$-100$~mV.

For a membrane voltage of $0$~mV, the energy maps of both helical
models reveal a trough bounded on all sides by regions of
substantially higher energy. The energy trough runs in the direction
of proportional translation and rotation for the \textalpha-helical S4
model, but it runs in the direction of simple translation for the
$3_{10}$-helical S4 model. The trough in the map of each model follows
the countercharge arrangement --- an arrangement chosen for each
model to allow periodic interactions during S4 motion of the S4
charges with the countercharges. In the \textalpha-helical model, the
S4 charges and countercharges are aligned on parallel super-helices,
whereas in the $3_{10}$-helical model the S4 charges and
countercharges follow straight lines parallel to the helical
axis. The lowest electrostatic energies of the \textalpha-helical S4
segment trace the path of a screw, whereas the energies of the
$3_{10}$-helical S4 segment trace the path of a piston.

The regions outside the energy trough for the $3_{10}$ helix have
energies about three times as large as those of the
\textalpha helix. The electrostatic confinement of the $3_{10}$ helix
is stronger than that of the \textalpha helix. The strength of
confinement correlates inversely with the separation of charges in the
two geometries. The cluster of S4 charges and countercharges is more
spread out in space in the \textalpha-helical than in the
$3_{10}$-helical geometry due to the angular separations of charges in
the \textalpha-helical S4 segment.

The electrostatic energy trough tends to anchor the sliding-helix in a
transmembrane configuration. The S4 charges that, for a given
configuration, dwell in the region of small polarizability are
balanced in this model by countercharges located in that region. This
balance is maintained over the range of S4 travel where equivalent
amounts of S4 charge and countercharge overlap in the region of weak
dielectric (see the Supplemental Material, animation
1~\cite{peyser:2012:supp:1} \& Fig.~\ref{fig:movie:standard} below). A
second essential element of balance concerns the transit of S4 charges
between the less polarizable gating canal region and the more
polarizable vestibule and bath regions. Any energy change associated
with the transit of an S4 charge on one side is approximately balanced
on the other side by the opposite transit of an S4 charge. On the
other hand, the energy trough generated by the electrostatics of the
models is too shallow by itself to ensure long-term stability of the
S4 configuration. Interactions beyond those included in the model
(such as linkages to adjacent transmembrane segments or hydrophobic
effect of the uncharged S4 residues) are necessary for long-term
stability.

To inspect the energetics more closely, we construct a one-dimensional
energy profile for S4 translation by computing for each translational
position the expectation of the electrostatic energy in the rotational
degree of freedom using the rotational partition function
[Eq.~(\ref{eq:randvarrot})]. We refer to this kind of energy profile
as a `translational energy profile' for
short. Figures~\ref{fig:Ebothhelices}(E) \& \ref{fig:Ebothhelices}(F)
show the translational energy profiles for two applied voltages:
$0$~mV (\emph{solid} lines) and $-100$ mV (\emph{dashed} lines). The
profiles at $0$~mV are quite uniform over the translational extent of
the energy trough (a consequence of the chosen spacing of
countercharges). At a membrane voltage of $-100$~mV, the energy
profiles are tilted in favor of more intracellular positions. A
well-defined energy minimum is found at a position about $-1$~nm
inward from the central position of the \textalpha-helical S4 segment,
whereas a broad minimum spread between $-0.7$ and $-1.2$~nm is found
with the $3_{10}$-helical S4 segment.

\begin{figure}
  \includegraphics{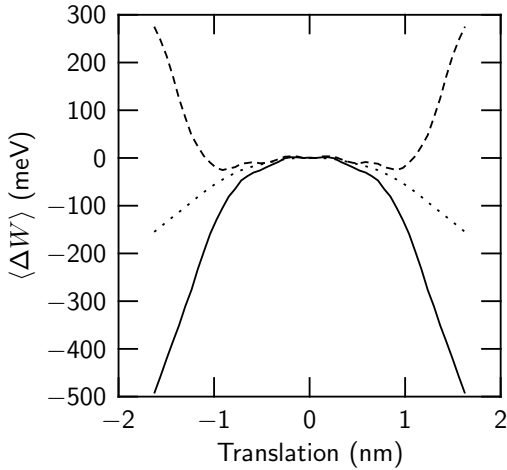}
  \caption{\emph{Energetic consequences of the countercharges.}
    Translational energy profile of the \textalpha-helical model of
    Fig.~\ref{fig:qvsbothhelices}(A) with countercharges present
    (dashed line) or deleted (solid line);
    $\epsilon_{\mathrm{p}}=4$. The dotted curve is computed with
    countercharges deleted and $\epsilon_{\mathrm{p}}=16$.}
  \label{fig:nocounter}
\end{figure}

The energy profiles in Fig.~\ref{fig:Ebothhelices} do not resemble the
profile of an ion embedded in a lipid membrane --- the latter profile
has a high barrier in the center of the weak dielectric
\citep{parsegian:1969,neumcke:1969}. Instead, in these models the
charged section of the S4 helix can travel with an almost level energy
over the translation range where S4 charges overlap with stationary
countercharges. To examine the contribution of the countercharges to
this result, we re-compute the energy profile for the
\textalpha-helical S4 segment with the countercharges deleted from
the model (Fig.~\ref{fig:nocounter}). Deletion of the countercharges
converts the energy trough seen with the standard model (\emph{dashed
  line}) into a broad barrier (\emph{solid line}). That barrier is
reduced but not inverted by increasing the VS dielectric coefficient
$\epsilon_{\mathrm{p}}$ from $4$ to $16$ (\emph{dotted line}). For
comparison, a sliding-helix model without countercharges using
$\epsilon_{\mathrm{p}}=10$ has been analyzed previously by
\citet{grabe:2004}.

Over the range of translation where the countercharges produce an
energy trough (\emph{dashed line}), the deletion of the
countercharges (\emph{solid line}) has a rather small effect (note
that we plot energy relative to the central position). Movement of the
S4 helix encounters little energy variation in this region of
translation because the amount of S4 charge present in the domain of
weak dielectric does not vary: as S4 charge enters on one side, S4
charge leaves on the other side. The energetics are not favorable for
these S4 positions (as the charges are not balanced), but they are
rather uniformly unfavorable until S4 translation exceeds \textsim1~nm
from its center position. If the travel of the S4 helix is restricted
to $\pm$$1$~nm in a biological channel by means other than the
countercharges deleted here, then even an S4 segment without
balancing countercharges could perform as a voltage sensor.

The variations of energy are small over the traveled range of
translation in Figs.~\ref{fig:Ebothhelices}(E) \&
\ref{fig:Ebothhelices}(F). The restriction in the model that the S4
domain and its charges must move as a single solid body might be
expected to `synchronize' periodic interactions among charges and
countercharges, leading the energetics to express several distinct
barriers and wells. The chosen spacing of the countercharges, however,
is enough to prevent the emergence of such a pattern. Additional
degrees of freedom are thus not a prerequisite for smooth S4 travel
(examples of such degrees of freedom are the possible flexibility of
the individual charge-bearing S4 residues or changes in configuration
of the helix between the $\alpha$ and $3_{10}$ forms,
\citealp{long:2007,khalili-araghi:2010,bjelkmar:2009,schwaiger:2011}).

Although the energy profiles of the two helix forms are similar, the
small differences between them are sufficient to produce substantial
differences in the relations between displaced charge and voltage
(Fig.~\ref{fig:qvsbothhelices}). Because of the small size of these
energy differences, contributions to the Hamiltonian not included in
the model (in particular contributions arising by coupling of the VS
to other parts of the channel) could override the differences between
the S4 helix versions seen in Fig.~\ref{fig:qvsbothhelices} and
Figs.~\ref{fig:Ebothhelices}(E) \& \ref{fig:Ebothhelices}(F).

Additional axes of variation for these models with consequences for
voltage gating include the geometry of the gating pore region, the
dielectric coefficient of the protein region, the distribution of
countercharges, and the effects of additional surface charges. We have
partially explored these spaces \citep{peyser:2011}. The robustness of
the mechanism explored here in the face of pathology due to mutation
can also be explored with this approach, given the computational
tractability of these mesoscale models.

\subsection{Energetics of  `working' voltage sensors}\label{sec:working}

\begin{figure*}
  \makebox[\textwidth]{%
    \hspace*{\stretch{1}}%
    \includegraphics{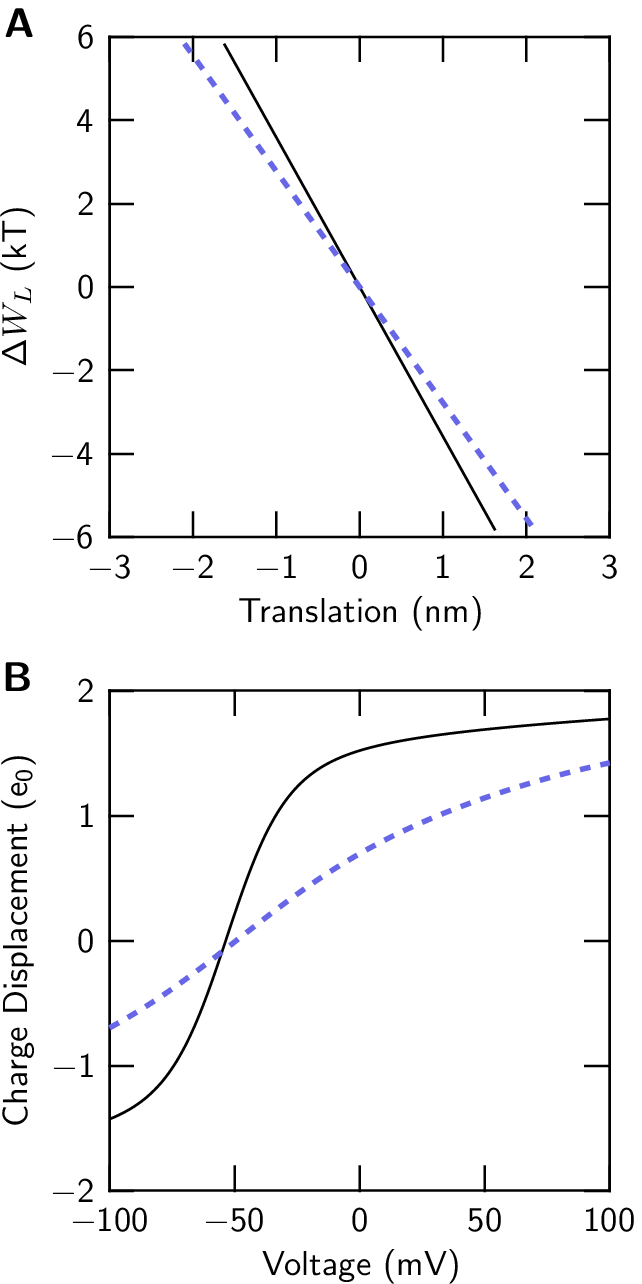}%
    \hspace*{\columnsep}%
    \includegraphics{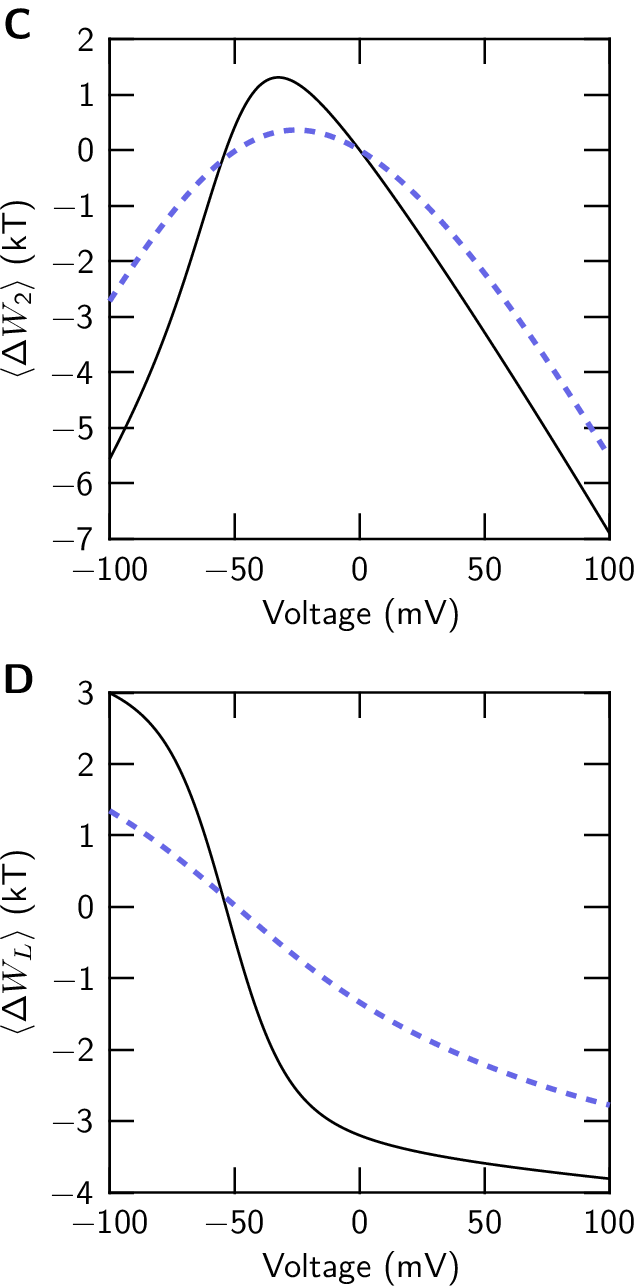}%
    \hspace*{\stretch{1}}%
  }
  \caption{\emph{Simulations of voltage sensors under a hypothetical
      load.}  (A) Component of the Hamiltonian representing the load
    [see Eq.~(\ref{eq:hamload})].  Expectations of displaced charge (B),
    the electrostatic energy due to the applied voltage (C), and the
    energy due to the interaction with the load (D). \emph{Solid
      lines:} $\alpha$-helical model; \emph{dashed lines:}
    $3_{10}$-helical model. All expectations of change in work are
    relative to the work for the configuration $(z, \phi) = (0, 0)$
    (which is identical for all $V_{\mathrm{m}}$, see
    Sec.~\ref{sec:stats}).}
    \label{fig:underload}
\end{figure*}

The essence of a voltage sensor is that it can do external work when
the membrane voltage is varied. This work can then be applied, for
instance, to reconfiguring the channel between conducting and
non-conducting configurations (gating). In order to determine the work
that a voltage sensor model might produce, we simulate models with a
translation-dependent external workload included in the
Hamiltonian. The potential energy field in which the S4 domain
translates and/or rotates then comprises the electrostatic potential energy
described by Eq.~(\ref{eq:ework}) plus the potential energy $W_L$ due to
the load:
\begin{equation}
W = W_1+W_2+W_L.
\label{eq:hamload}
\end{equation}
For our examples, we use a hypothetical load field producing a
constant force opposing the inward translation of the S4 segment and
hence a load potential that varies linearly with translation
[Fig.~\ref{fig:underload}(A)]. This describes the energetics if, for
instance, the gate of a hypothetical channel with a single VS resists
closing (by inward movement of the S4 helix) with a constant force.

The presence of this load alters the relation between the mean
displaced charge and voltage by shifting the curve to negative
voltages [Fig.~\ref{fig:underload}(B), compare
Fig.~\ref{fig:qvsbothhelices}]. In the case of the \textalpha-helical
S4 model, the charge-voltage curve becomes quite similar to that
observed in the experiment with full channels [symbols in
Fig.~\ref{fig:qvsbothhelices}(A)].

In order to assess the relationship between the electrical work picked
up by the VS and the external work delivered to the load, we consider
the expectations of $\Delta W_2$ [Fig.~\ref{fig:underload}(C)] and
$\Delta W_L$ [Fig.~\ref{fig:underload}(D)]. The expectation for $\Delta
W_2$ has a simple relationship to the expectation of displaced charge:

\begin{equation}
\langle\Delta W_2\rangle = - \sum_{i,j} Q_{ij}V_{\mathrm{m}}P_{ij}
  = - \langle Q\rangle V_{\mathrm{m}}
\label{eq:workin}
\end{equation}
 
The work exchanged between the VS and the load ($\langle\Delta
W_L\rangle$) varies most strongly with voltage where the variation of
displaced charge is maximal, as expected if the load depends on S4
position. Two points on the axis of applied voltage are associated
with polarity changes of displaced charge and/or of components of the
Hamiltonian in Fig.~\ref{fig:underload} (not marked): $V_1$ for
$\langle Q\rangle$ (B), $\langle\Delta W_2\rangle$ (C), and
$\langle\Delta W_L\rangle$ (D); and $V_2=0$ for $\langle\Delta
W_2\rangle$ (C). In the voltage region $V_{\mathrm{m}}<V_1$, the
applied voltage drives the S4 helix inward while external work is done
against the load. In the region $V_{\mathrm{m}}>V_2$, both the applied
voltage and the work done on the S4 helix by the load drive the S4 helix
outward ($\langle Q\rangle>0$). In the intermediate region $V_1<
V_{\mathrm{m}}<V_2$, the S4 helix is also driven outward ($\langle
Q\rangle>0$). Here the applied-voltage and load components of the
Hamiltonian have opposite polarities: outward S4 motion prevails
because the load prevails over the opposing effect of applied voltage.

The translation range traveled by the S4 helix is limited in these
simulations by the electrostatics of the VS rather than by the
hypothesized load. The limits are set by the electrostatic self-energy
component of the Hamiltonian, $\Delta W_1$
(Fig.~\ref{fig:Ebothhelices}, $V_{\mathrm{m}}=0$). Therefore, the work
that the VS can do on the load at large (positive or negative)
voltages saturates [Fig.~\ref{fig:underload}(D)], whereas the (negative)
energy contributed to the VS by the applied field continues to
increase in magnitude [Fig.~\ref{fig:underload}(C)]. Biological K$^+$
channels open with very low probability ($<10^{-6}$) at large negative
voltages with no indication of a saturating minimal probability
\citep{seoh:1996,schoppa:1992,aggarwal:1996}. This may indicate that
the load imposed on the VS domains by the `gate' of the channel
actually limits S4 travel at large negative voltages. In contrast, at
large positive voltages the open probability of the channels saturates
at levels well below $1$, suggesting a saturating amount of work that
the VS domains can do on the gate.

Figure~\ref{fig:underload} shows the simulation results for both the
$\alpha$ and $3_{10}$ configurations of the S4 helix. The differences
in voltage responsiveness observed in the simulations with the idling
sensors are also found under the hypothetical load that we test. This
observation, however, cannot be generalized. The differences between
these two forms of VS result from the associated electrostatic
self-energies $\Delta W_1$ (as seen in
Fig.~\ref{fig:Ebothhelices}). In the Hamiltonian of the system under
load [Eq.~(\ref{eq:hamload})], characteristics are determined by the sum
of the self-energy term $W_1$ and the load term $W_L$. Hence, it is in
principle possible that the load transforms the characteristics of
voltage sensing observed here for the $3_{10}$-helical VS model into
characteristics indistinguishable from those of the $\alpha$-helical
VS model or vice-versa, or a load may even result in behaviors
different from those seen in either simulation. Experimental
observations on displaced charge therefore cannot be interpreted in
terms of models that include only a voltage sensor. These
observations, like those on gating, require a model of the full
channel to be analyzed.

The mesoscale model system that we have presented for a single VS
is extendable using the approach described in this section. The Hamiltonian of
the system can be extended by terms describing the energetics of a
channel comprising four voltage sensors, a gating domain, and their
coupling. The number of degrees of freedom (e.g., two for each voltage
sensor) remains computationally manageable, so that a mesoscale
model of the full channel can be simulated in order to obtain insight
into the cooperation of its parts.

We have analyzed here some equilibrium properties of a VS domain in
which the S4 helix is relocated as a solid body. A refined model
augmented with dynamics will likely have to include a second scale
describing motions of side chains, possible deformations of the S4
helix, interactions with the bath solutions, and associated
dissipative aspects. General energetic variational approaches
developed for hydrodynamic systems of complex fluids \citep{hyon:2010}
may provide a multi-scale approach to channel dynamics.

\begin{acknowledgments}
  The authors are grateful for the support of the National Institutes
  of Health (Grant No.~GM083161) to W.N.~and a Graduate Research
  Fellowship of the National Science Foundation to A.P. We thank
  Dr.~Alice Holohean, Dr.~Peter Larsson, and Dr.~Karl Magleby for
  helpful discussions.
\end{acknowledgments}

\vskip\baselineskip
\appendix*
\section{Charge distributions of VS models}

\begin{figure*}
  \includegraphics[width=\linewidth]{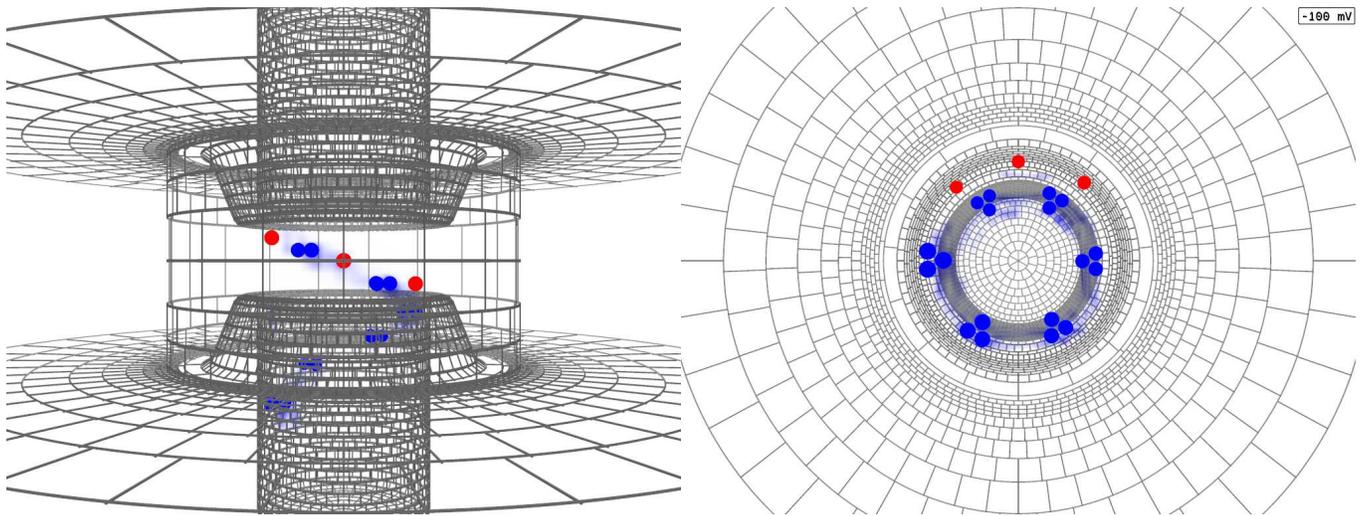}
  \caption{\emph{Standard \emph{\textalpha}-helical
      model:} Position and distribution of charges for model
    \textalpha~(1) in Table~\ref{fig:cell}(C) with $\epsilon_p=4$. Red
    (light gray) symbols represent fixed negative countercharges,
    blue (dark gray) symbols represent the mean position of
    $-1/3$~$\text{e}_0$ on S4 arginines, and blue (gray) shading
    represent the relative probability of negative charge at a given
    position. This is the model used in
    Fig.~\ref{fig:qvsbothhelices}(A). See the associated Supplementary
    Material, animation~1~\cite{peyser:2012:supp:1} for the behavior of this
    model of the VS domain over the range of transmembrane potentials
    from $-100$~mV to $+100$~mV.}
    \label{fig:movie:standard}
    \label{fig:movie:first}
\end{figure*}

\begin{figure*}
  \includegraphics[width=\linewidth]{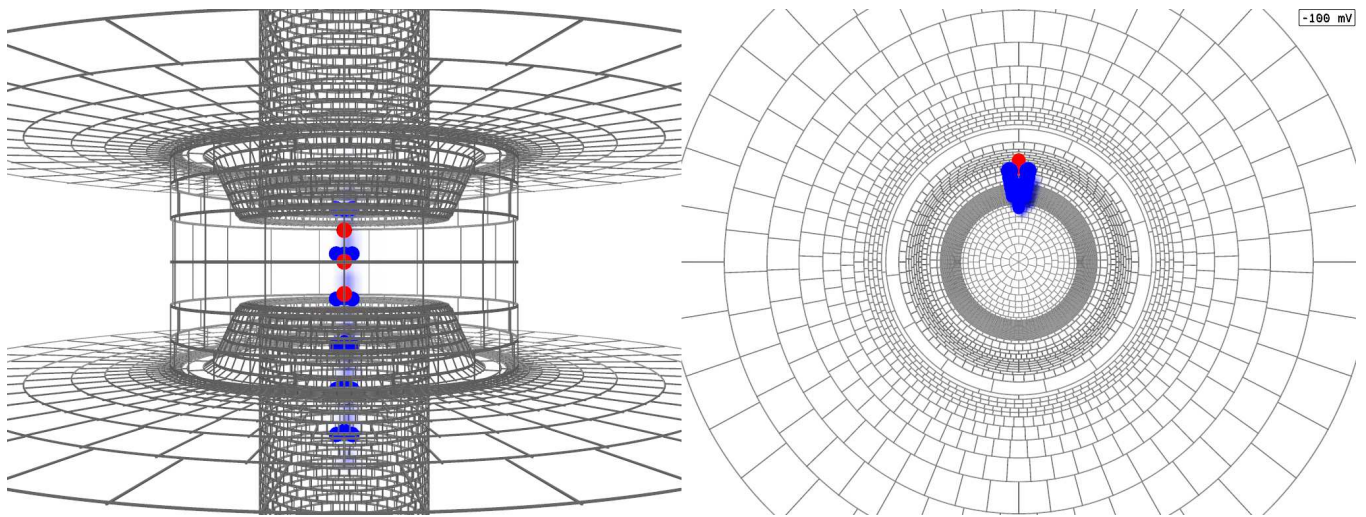}
  \caption{\emph{Standard $3_{10}$-helical model:}
    Position and distribution of charges for model $3_{10}$ in
    Table~\ref{fig:cell}(C) with $\epsilon_p=4$. See the description of
    Fig.~\ref{fig:movie:first} for further details.  This is the model
    used in Fig.~\ref{fig:qvsbothhelices}(B). See the associated Supplemental
    Material, animation~2~\cite{peyser:2012:supp:2} for the behavior of this
    model of the VS domain over the range of transmembrane potentials
    from $-100$~mV to $+100$~mV.}
  \label{fig:movie:310}
  \label{fig:movie:last}
\end{figure*}

The sliding helix in our VS models is a microscopic voltage
sensor. Therefore, we describe the VS by its statistical ensemble
behavior. In this appendix, stochastic VS behavior is visualized for
several models presented in the paper, both as figures for a fixed
applied voltage ($-100$~mV; Figs.~\ref{fig:movie:first} \&
\ref{fig:movie:last}) and as associated Supplemental Materials,
animations~1~\cite{peyser:2012:supp:1} \& 2~\cite{peyser:2012:supp:2}
with voltage increasing uniformly over time from $-100$~mV to
$+100$~mV.

The figures show two stochastic aspects of VS behavior: (1) the mean
positions of the S4 charges [marked by blue (dark gray) balls], and
(2) the charge density distribution of S4 charge [represented by a
blue (gray) cloud with a color intensity proportional to the charge density
there]. A high density of color marks the locations where the S4
charges dwell frequently, as opposed to their mean positions.

The expectation of position for each charge is computed from
Eq.~(\ref{eq:randvar}) using the positions $\mathbf{r}_k$ for the charge
$q_k$ as the random variable $X$, over the partition function with
translational and rotational degrees of freedom.  Since the helix
behaves as a solid body, the helix position $\mathbf{r}$ fixes the
positions $\mathbf{r}_k$ for the charges $q_k$. That relationship
allows us to define the partition function, the energy function, the
configuration probability and our measures for the positions
$\mathbf{r}_k$ in terms of the respective function for the helix
position $\mathbf{r}$. For the expectation of position for each
charge:
\begin{equation}
  \langle\mathbf{r}_k\rangle = \sum_{i,j} \mathbf{r}_{ijk}P_{ij}
  = \frac{1}{\mathcal{Q}} \sum_{i,j} \mathbf{r}_{ijk} e^{-W_{ij}/k_BT},
  \label{eq:pos:mean}
\end{equation}
where $W_{ij}$ is the work needed to construct configuration $ij$, $P_{ij}$
is the probability of that configuration, and
$\mathbf{r}_{ijk}$ is the position of charge $k$ in configuration
$ij$.

Likewise, the distribution of charge can be computed by applying the
partition and energy functions in terms of the positions
$\mathbf{r}_k$ of charges $q_k$. The charge density $\bar
z(\mathbf{r})$ is then the sum over all charges of the probability of
each charge being located at $\mathbf{r}$, multiplied by its valency
in units of $\mathrm{e}_0$, and normalized:
\begin{equation}
  \bar z(\mathbf{r}) 
  = \Bigl[\sum\limits_{i,j,k} P_{ij}z_k\Bigr]^{-1}
    \sum_{i,j,k} P_{ij}z_k\,\delta(\mathbf{r},\mathbf{r}_{ijk}),
  \label{eq:pos:prob}
\end{equation}
where $\delta(\mathbf{r}, \mathbf{r}_{ijk})$ is the discretized delta
function ($1$ if we are treating $\mathbf{r}_{ijk}$ as the same
location as $\mathbf{r}$ for visualization purposes; otherwise
$0$).

The color representations for the animations are proportional to $\bar
z(\mathbf{r})$, normalized to the highest charge density at that
frame's potential.

\bibliography{pubmed-abbrv,other-abbrv,tentative,all}

\end{document}